\documentclass[%
reprint,
superscriptaddress,
showpacs,
preprintnumbers,
citeautoscript,
amsmath,
amssymb,
aps,
prb,
]{revtex4-1}

\usepackage{graphicx}
\usepackage{dcolumn}
\usepackage{bm}
\usepackage{xcolor}
\usepackage{comment}
\usepackage[colorlinks,urlcolor=blue,citecolor=blue]{hyperref}
\usepackage{subfigure}
\usepackage{mathtools}
\usepackage{comment}

\usepackage{bm}
\renewcommand {\vec} [1] {{\bm #1}}

\begin{document}

\title{Electron dynamics in extended systems within real-time time-dependent density functional theory}

\author{Alina Kononov}
\affiliation{Center for Computing Research, Sandia National Laboratories, Albuquerque, NM 87123, USA}

\author{Cheng-Wei Lee}
\affiliation{Colorado School of Mines, Golden, CO 80401, USA}

\author{Tatiane Pereira dos Santos}
\affiliation{Department of Materials Science and Engineering, University of Illinois at Urbana-Champaign, Urbana, IL 61801, USA}

\author{Brian Robinson}
\affiliation{Department of Materials Science and Engineering, University of Illinois at Urbana-Champaign, Urbana, IL 61801, USA}

\author{Yifan Yao}
\affiliation{Department of Materials Science and Engineering, University of Illinois at Urbana-Champaign, Urbana, IL 61801, USA}

\author{Yi Yao}
\affiliation{Thomas Lord Department of Mechanical Engineering and Materials Science, Duke University, Durham, NC, 27708, USA}
\affiliation{Department of Chemistry, University of North Carolina at Chapel Hill, NC, 27595, USA}

\author{Xavier Andrade}
\affiliation{Quantum Simulations Group, Lawrence Livermore National Laboratory, Livermore, California 94551, United States}

\author{Andrew David Baczewski}
\affiliation{Center for Computing Research, Sandia National Laboratories, Albuquerque, NM 87123, USA}

\author{Emil Constantinescu}
\affiliation{Mathematics and Computer Science, Argonne National Laboratory, Lemont, IL 60439, USA}

\author{Alfredo A.\ Correa}
\affiliation{Quantum Simulations Group, Lawrence Livermore National Laboratory, Livermore, California 94551, United States}

\author{Yosuke Kanai}
\affiliation{Department of Chemistry, University of North Carolina at Chapel Hill, NC, 27595, USA}

\author{Normand Modine}
\affiliation{Material, Physical, and Chemical Sciences Center, Sandia National Laboratories, Albuquerque, NM 87123, USA}

\author{Andr\'e Schleife}
\email{schleife@illinois.edu}
\affiliation{Department of Materials Science and Engineering, University of Illinois at Urbana-Champaign, Urbana, IL 61801, USA}
\affiliation{Materials Research Laboratory, University of Illinois at Urbana-Champaign, Urbana, IL 61801, USA}
\affiliation{National Center for Supercomputing Applications, University of Illinois at Urbana-Champaign, Urbana, IL 61801, USA}

\date{\today}

\begin{abstract}
Due to a beneficial balance of computational cost and accuracy, real-time time-dependent density functional theory has emerged as a promising first-principles framework to describe electron real-time dynamics.
Here we discuss recent implementations around this approach, in particular in the context of complex, extended systems.
Results include an analysis of the computational cost associated with numerical propagation and when using absorbing boundary conditions.
We extensively explore the shortcomings for describing electron-electron scattering in real time and compare to many-body perturbation theory.
Modern improvements of the description of exchange and correlation are reviewed.
In this work, we specifically focus on the Qb@ll code, which we have mainly used for these types of simulations over the last years, and we conclude by pointing to further progress needed going forward.
\end{abstract}

\maketitle

\section{\label{sec:intro}Introduction}

Real-time time-dependent density functional theory (RT-TDDFT) has attracted tremendous attention in the context of accurate theoretical characterization of materials recently and over the years.
It is arguably one of the most promising approaches to simulate the real-time quantum dynamics of electrons as well as its coupling to ion dynamics.
In particular, its promising balance between accuracy and computational cost make this technique increasingly applicable also for development, design, and discovery of materials including for electronic, optical, electrochemical applications, amongst others \cite{santos2021}.
Recent applications include laser excitation of materials \cite{Miyamoto:2021}, interaction of materials with energetic ions \cite{Correa:2018}, and nonlinear excitation dynamics \cite{Uemoto:2019}.
The framework is implemented in many software packages and readily usable on a large variety of computational resources, including use of graphics processing units (GPUs).
This makes the technique applicable to many diverse materials from just a few atoms to complex extended structures consisting of hundreds of atoms.

In this work we provide examples for recent developments and applications that we accomplished and use these to illustrate the need for future improvements.
This includes discussing the underlying approximations and the path towards a computationally more feasible and widely applicable implementation of this approach for complex and extended systems.

First, the time stepping that is used in RT-TDDFT critically determines the computational cost.
Second, we also give a specific example for how  absorbing boundary conditions can mitigate high computational cost when studying two-dimensional materials.
Next, the physics of charged projectile ions or electrons interacting with the electronic system of the target is briefly discussed and the computational cost of using an electron wave packet instead of a classical Coulomb potential in a plane-wave framework is assessed.
Subsequently, we analyze in detail the RT-TDDFT description of electron dynamics and find shortcomings in capturing the time scale of electron-electron scattering mediated thermalization.
These results are compared to the literature and discussed relative to \emph{GW} simulations within many-body perturbation theory.
Finally, we discuss recent progress in describing the electron-electron interaction via exchange and correlation in RT-TDDFT, and the associated computational cost.
All RT-TDDFT simulations presented here were performed with the Qb@ll code and extensions thereof \cite{schleife:2012,Schleife:2014,draeger:2017,qball,gygi:2008}, and we conclude our discussion with a brief outlook on future directions of this software, hoping to stimulate exciting developments in the field of RT-TDDFT for years to come, including for computational materials discovery and development, as is the goal of this focus issue.

\section{Real-time propagation of time-dependent Kohn-Sham equations}

Excited electron dynamics can be modeled from first principles with real-time time-dependent density functional theory (TDDFT) \cite{runge:1984,marques:2004,ullrich:2012,ullrich:2014}.
In this approach, the electron density $n(\mathbf{r},t)$ evolves over time according to the time-dependent Kohn-Sham (TDKS) equations:
\begin{equation}
    \begin{aligned}
    i\frac{\partial}{\partial t} \phi_j(\mathbf{r},t) = \hat{H}[n](t)\,\phi_j(\mathbf{r},t),\\
n(\mathbf{r},t)=\sum_j f_j\left|\phi_j(\mathbf{r},t)\right|^2.
    \end{aligned}
    \label{eq:tdks}
\end{equation}
Here, $\phi_j$ are single-particle Kohn-Sham orbitals with occupations $f_j$.
The single-particle Hamiltonian,
\begin{equation}
    \label{eq:hamiltonian}
    \hat{H}[n](t)=
    \hat{T} + \hat{V}_{\mathrm{ext}}(t) + \hat{V}_{\mathrm{Har}}[n] + \hat{V}_{\mathrm{XC}}[n],
\end{equation}
contains the kinetic energy operator $\hat{T}$, the external potential $\hat{V}_{\mathrm{ext}}(t)$ due to nuclei and any external fields, the Hartree electron-electron potential $\hat{V}_{\mathrm{Har}}[n]$, and the exchange-correlation potential $\hat{V}_{\mathrm{XC}}[n]$.
The electronic system may be coupled to nuclear motion through Ehrenfest dynamics \cite{ehrenfest:1927,Andrade2009}.

Explicit time dependence may arise within $\hat{V}_{\mathrm{ext}}(t)$ from an external perturbation such as a moving projectile ion or a dynamic electromagnetic field.
Depending on the gauge choice, an external vector potential $\mathbf{A}_{\mathrm{ext}}(\mathbf{r},t)$ may enter into the kinetic energy as \mbox{$\hat{T}=\frac{1}{2}\left(-i\nabla + \mathbf{A}_{\mathrm{ext}}(\mathbf{r},t)\right)^2$}.
To apply a uniform external electric field to an infinite periodic system, it is often convenient to work in the velocity gauge, where the electric field is generated by the vector potential ~\cite{Bertsch2000,yabana:2006,krieger:2015,andrade:2018,sun:2021} \begin{equation}
\label{eq:egauge}
    \mathbf{E}_{\mathrm{ext}}(t) = -\frac{1}{c}\frac{d\mathbf{A}_{\mathrm{ext}}(t)}{dt} \ .
\end{equation} 
Alternatively, the length gauge, which instead involves the scalar potential $\mathbf{E}_{\mathrm{ext}}(t)\cdot \mathbf{r}$, can be appropriate for finite systems \cite{tsolakidis:2002,takimoto:2007} or with the use of maximally localized Wannier functions \cite{yost:2019}.
Both capabilities have been implemented in the plane-wave TDDFT code Qbox/Qb@ll \cite{schleife:2012,Schleife:2014,draeger:2017}, with options for static fields, delta kicks, and dynamic laser pulses~\cite{yost:2019,sun:2021}.

In the vector-potential formulation, the vector potential is chosen such that its time derivative gives the proper electric field according to Eq.\ \eqref{eq:egauge}.
For example, the delta kick is implemented by a step function in the vector potential.
In practice this means the propagation is done with a constant vector potential whose amplitude is given by a desired intensity of the kick (as the initial condition is the ground state calculated without a vector potential).
A laser field is simply simulated by an oscillatory electric field with constant or time-dependent amplitude.
Since the dipole is not properly defined for extended systems, the polarization is obtained from the macroscopic current \cite{Bertsch2000}.
We use the usual definition of the quantum-mechanical current 
\begin{equation}
\label{eq:current}
J(t)=\int\mathrm{d}\mathbf{r}\sum_j f_j\phi^*_j(\mathbf{r},t)\vec{\nabla}\phi_j(\mathbf{r},t) + c.c.\ ,
\end{equation}
which it is not strictly correct when using non-local pseudo-potentials \cite{andrade:2018}.
However the correction term is small for electric perturbations \cite{Luppi2008}.

Both the computational cost and accuracy of real-time TDDFT simulations are in large part governed by the numerical algorithm used to integrate the TDKS equations, Eq.\ \eqref{eq:tdks}.
While a simple explicit integration scheme such as fourth-order Runge-Kutta (RK4) is suitable for modest-size systems~\cite{schleife:2012}, very large supercells  and short-time propagation require higher accuracy \cite{Kang:2019} offered by time-reversible schemes such as the enforced time-reversal symmetry (ETRS) method \cite{castro:2004,draeger:2017}.
We specifically showed this for systems containing vacuum \cite{kononov:2020}.
More efficient algorithms which reduce time-to-solution without sacrificing accuracy would accelerate the study of excited electron dynamics in materials and enable consideration of larger systems of practical interest over longer simulation time scales, including defect systems, material surfaces, and 2D hetero-structures.

Below we briefly present our recent efforts towards a systematic assessment of numerous explicit time-steppers and several variants of the ETRS approach.
Interfacing Qbox/Qb@ll \cite{schleife:2012,draeger:2017} with the PETSc numerical library \cite{balay:1997,abhyankar:2018,balay:2019:web,balay:2020:user-ref} provided us with seamless access to a wide range of Runge-Kutta (RK) \cite{bogacki:1989,bogacki:1996,dormand:1980,fehlberg:1970} and strong stability preserving (SSP) RK  \cite{kraaijevanger:1991,ruuth:2006,ketcheson:2008,gottlieb:2009,gottlieb:2015} methods.
Each algorithm's performance was assessed for a sodium dimer test system over a range of time step sizes $\Delta t=0.01$\,--\,0.5 atomic units (at.\,u.), and computational cost was measured as the average wall time per simulated time.
After perturbing the initially ground-state system by slightly displacing the atoms away from their equilibrium positions, the electronic response was evolved for 100 time steps on a single processor.
For the most promising methods, additional tests on a 112-atom graphene supercell confirmed the qualitative trends observed for the smaller test system.

Since exact time evolution should conserve both energy and charge, we compute an error metric given by the product of average errors in total energy $E$ and net charge $Q$ per simulation time:
\begin{equation}
\label{eq:error_metric}
\delta Q\, \delta E = \left\langle \frac{Q(t)-Q(0)}{t} \right\rangle \left\langle \frac{E(t)-E(0)}{t} \right\rangle ,
\end{equation}
where brackets denote time averages,
\begin{equation}
\label{eq:timeaverage}
\left\langle \frac{Q(t)-Q(0)}{t} \right\rangle = \frac{1}{t_f}\int_0^{t_f} \frac{Q(t)-Q(0)}{t} \; dt,
\end{equation}
and $t_f$ is the total time.
This form was chosen to give a reasonable error measure for both linear and oscillatory error accumulation models.
In particular, using this definition $\delta Q$ has a reasonable long-time limit both when $Q(t)-Q(0)$ can be modeled as $\propto t$ and when $Q(t)-Q(0)$ can be modeled as $\propto \sin(\omega t)$ \cite{kononov_thesis}.
The tolerable error level for a particular application depends on the system studied and the observable of interest.
For example, electronic stopping power calculations in bulk materials \cite{pruneda:2007,schleife:2015,magyar:2016,lim:2016,yost:2016,yost:2017,lee:2018,ullah:2018,lee:2019,halliday:2019,lee:2020,qi:2022} extract total energy differences typically about 5\,--\,50\,Ha over the course of a $\sim$1\,fs simulation, so $\delta E \ll 0.1$\,Ha/at.\,u.\ suffices and $\delta Q$ is not important beyond its correlation with $\delta E$.
In contrast, simulations of ion-irradiated 2D materials \cite{zhang:2012,ojanpera:2014,zhao:2015,kononov:2020,kononov_anomalous_2021,vazquez:2021,kononov:2022} involve smaller energy transfers around 0.2\,--\,5\,Ha 
and may additionally examine sensitive charge transfer processes such as emission of 0.1\,--\,10 electrons into vacuum.
Thus, these calculations require $\delta Q\, \delta E \ll 10^{-5}$\,e\,Ha/at.\,u.$^2$

\begin{figure}
\centering
\includegraphics[width=\columnwidth]{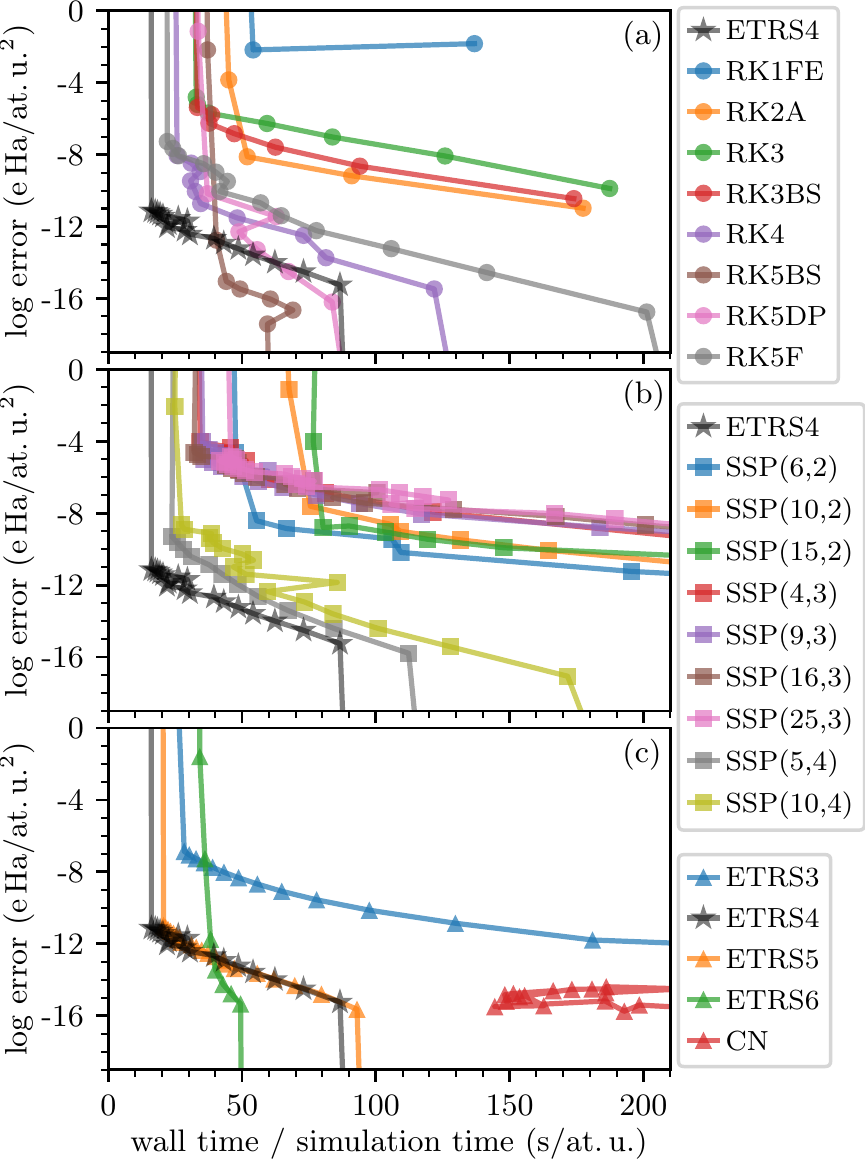}
\caption{\label{fig:integrators}
Performance of 4th-order ETRS (black stars) compared to (a) all Runge-Kutta time-steppers available in PETSc, (b) various strong stability preserving Runge-Kutta time-steppers available in PETSc, and (c) other variants of ETRS and a naive application of PETSc's Crank-Nicolson (CN).
RK$N$[X] denotes an $N$th-order Runge-Kutta scheme where X is an additional PETSc identifier, typically the initials of original developers.
SSP$(M,N)$ denotes an $M$-stage, $N$th-order SSPRK method, and ETRS$N$ denotes ETRS using $N$th-order Taylor expansions to approximate exponentials.
}
\end{figure}

From our data in Fig.\ \ref{fig:integrators} we find that ETRS generally outperforms all explicit time steppers tested: it achieves lower computational cost at an acceptable error level.
The only competitive Runge-Kutta scheme is the fifth-order Bogacki-Shampine algorithm (RK5BS) \cite{bogacki:1996}, which is even more accurate than ETRS for small step sizes (see Fig.\ \ref{fig:integrators}a).
However, while RK5BS becomes unstable for $\Delta t\gtrsim 0.1$\,at.\,u.\ in our sodium dimer simulations, ETRS maintains tolerable error rates for step sizes twice as large, allowing lower computational cost.
Among the SSP methods tested, the 4th-order schemes are most successful but do not improve over ETRS's accuracy, stability, or speed (see Fig.\ \ref{fig:integrators}b).
Lower-order SSP schemes involving many ($\geq 16$) stages do allow larger step sizes than ETRS, but the expense associated with a large number of stages outweighs the increased stability.
Overall, we find that ETRS achieves lowest time-to-solution.
Recent work \cite{rehn:2019} also tested the Adams-Bashforth and Adams-Bashforth-Moulton classes of explicit time steppers, finding that these methods can outperform RK under certain conditions, but their performance has not yet been compared to ETRS.

Several possible schemes exist to approximate the exponentials of the Hamiltonian involved in ETRS \cite{castro:2004}.
Here, we use Taylor expansions for their simplicity and compare different orders in Fig.\ \ref{fig:integrators}c.
Consistent with assertions made in Ref.\ \onlinecite{castro:2004}, we find that 4th or 5th-order Taylor expansions are optimal.
A 6th-order expansion is less stable, while a 3rd-order expansion sacrifices accuracy without significantly reducing computational cost.

Other implicit methods may yet prove more efficient than ETRS.
One promising option is Crank-Nicolson (CN), which some other TDDFT implementations successfully employ \cite{tsolakidis:2002,qian2006time,meng:2008,ojanpera2012nonadiabatic,baczewski:2014,baczewski2016x,magyar:2016}.
We find that CN is generally more accurate than ETRS (see Fig.\ \ref{fig:integrators}c), perhaps thanks to the unitarity of the Pad\'{e} form of the CN propagator in contrast to the truncated Taylor expansion used in the ETRS implementation.
Although CN can maintain accuracy even for large time steps, i.e., stability restrictions do not limit this method, it involves a costly nonlinear solve.
The large number of $\hat{H}\phi$ evaluations performed by PETSc's algorithm for this nonlinear solve made CN prohibitively expensive in this work (see Fig.\ \ref{fig:integrators}c).
However, further optimization, efficient preconditioners, or the use of predictor-corrector methods that obviate the nonlinear solve~\cite{qian2006time,walter2008time} could alleviate this issue.
Implicit schemes such as CN could be particularly advantageous for ultrasoft pseudopotentials~\cite{vanderbilt1990soft} or the projector augmented-wave method~\cite{blochl1994projector}, where the left-hand side of the TDKS equations involves an overlap matrix acting on the time derivative of the pseudized orbitals \cite{baczewski:2014}.
Since explicit time-stepping schemes require the application of the inverse of this matrix at each time step, this complication narrows the prospective efficiency gap between explicit and implicit schemes.
However, this work used norm-conserving pseudopotentials \cite{troullier:1991} and thus did not benefit from CN.

Explicit RK methods cannot conserve energy, and some of the least expensive implicit RK methods, such as CN, do not in general.
In general, a direct way to alleviate errors in invariant quantities represented by inner-product norms is to control the time step.
One can reduce the time step adaptively if the energy loss exceeds a certain level.
Moreover, a promising strategy that also applies to explicit methods is to use a time-step adaptation that adjusts step length such that the energy is conserved exactly in finite precision.
These methods are referred to as relaxation RK and rely on modifying the prescribed time step (typically reducing it by a small fraction) so that the solution at each of these modified steps preserves energy \cite{Kang_2021b,ketcheson2019relaxation}.
Explicit methods are conditionally stable; nevertheless, the stability regions can be optimized for a specific eigenvalue portrait, which is a promising strategy to improve their performance.
Furthermore, new machine learning developments in neural ODE may provide new ways to accelerate the time stepping process \cite{liang2022stiffnessaware}. 

Finally, the parallel transport gauge approach \cite{jia:2018,jia:2019,an:2020,an:2022} applies a unitary transformation to the Kohn-Sham orbitals to instead solve for slower varying orbitals that reproduce the same electron density but introduce an additional term in the TDKS equations.
This promising method can be combined with an efficient time stepper to produce speedups of 5\,--\,50 over standard RK4 for molecules \cite{jia:2018,jia:2019,an:2020}, solids containing up to 1024 atoms \cite{jia:2018,jia:2019}, and mixed states in model systems \cite{an:2022}.

\section{Complex Absorbing Potential for Secondary Electron Emission}

After examining the computational cost associated with real-time propagation in the previous section, we also explored the need for a large vacuum region as part of the simulation cell when studying electron emission, e.g.\ from surfaces or two-dimensional (2D) materials.
When using periodic boundary conditions, vacuum lengths of $150~a_0$ or more are necessary to prevent the unphysical interaction of the electrons emitted from both sides of the 2D material across the boundary of the simulation cell, resulting in a high computational demand \cite{kononov_anomalous_2021}.
To address this problem, absorbing boundaries\cite{de_giovannini_modeling_2015,wopperer_efficient_2017} are frequently employed to emulate open boundary conditions.
Absorbing boundaries based on a complex absorbing potential (CAP)\cite{muga_complex_2004} alter the Hamiltonian, Eq.~\eqref{eq:hamiltonian}, by adding an artificial complex (imaginary) potential in a defined region of the simulation cell, resulting in a non-Hermitian Hamiltonian and non-unitary time-evolution operator.
This approach has been successfully used in simulating the real time dynamics of wave functions of 2D materials, including secondary electron emission due to electron irradiation \cite{ueda_quantum_2016,ueda_time-dependent_2018,miyauchi_electron_2017} and angular resolved photo-emission spectra \cite{de_giovannini_first-principles_2017,wopperer_efficient_2017}.

We implemented an absorbing potential into the Qb@ll~\cite{schleife:2012,draeger:2017} code that follows the form
\begin{equation}
\label{eq:cap}
\mathrm{V}_{\mathrm{CAP}}(z)=
\begin{cases}
-i\cdot W \sin^2\left(\frac{(z-z_s)\cdot \pi}{2\cdot d_z}\right), & \text{$z_s<z<z_s+2 d_z$} \\
0, & \text{otherwise}
\end{cases}
\end{equation}
where $W$ defines the maximum of the CAP, and $z_s$ and $d_z$ are the position of the front boundary and the half width of the CAP.

\begin{figure}
\centering
\includegraphics[width=\columnwidth]{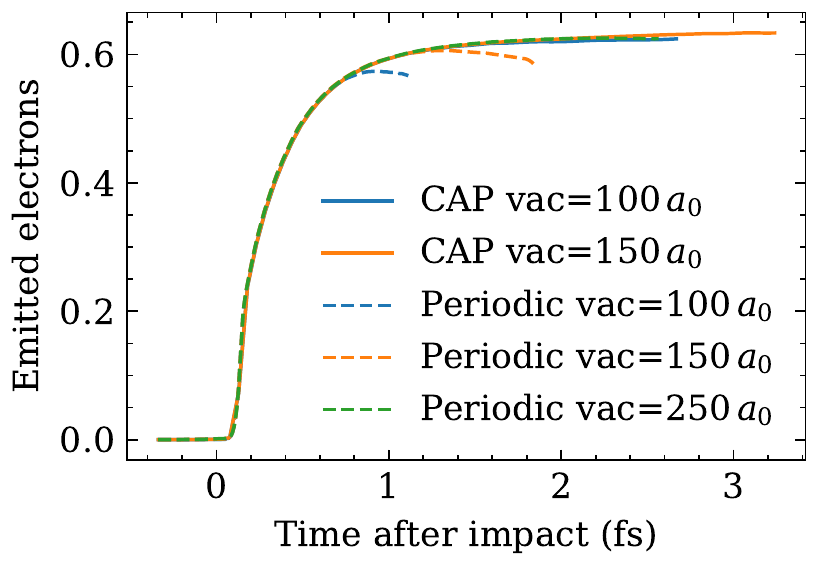}
\caption{\label{fig:vaccom}
Total emitted electrons in vacuum, after a channeling proton\cite{kononov_anomalous_2021} with a velocity of  1.79 at.\ u.\ impacts graphene.
When using a CAP, the difference between different vacuum sizes converges much earlier.
}
\end{figure}

Here, we compare to our previous work on secondary electron emission of graphene under proton irradiation \cite{kononov_anomalous_2021}, and demonstrate that a CAP can significantly reduce finite size effects, leading to an acceleration of the simulation by reducing the vacuum size.
We use the same simulation cell and computational parameters as described in Ref.\ \onlinecite{kononov_anomalous_2021}.
The target graphene is placed at the center of the simulation cell, at $z=0$ on the $x\mbox{-}y$ plane. Emitted electrons in vacuum are determined by integrating the electron density over a region farther than $10.5\,{a_0}$ from the graphene.
We assess finite size effects for different vacuum sizes along the direction of proton travel for a channeling proton with 1.79 at.\,u.\ of velocity.
Following Ref.\ \onlinecite{kononov_anomalous_2021}, we treat the maximum of the emitted electron curves in Fig.\ \ref{fig:vaccom} as the total number of emitted electrons.

Comparing the resulting number of total emitted electrons for periodic boundary conditions, the data in Fig.\ \ref{fig:vaccom} shows a difference of 3\,\% when 150\,${a_0}$ and 250\,${a_0}$ of vacuum are used, whereas the difference is 8.22\,\% between 100\,${a_0}$ and 250\,${a_0}$ of vacuum.
This shows that a large vacuum size is needed to obtain converged results.
For comparison, a CAP of the form of Eq.\ \eqref{eq:cap} is placed at the boundary of the simulation cell.
We set $W=15\,E_\mathrm{h}$, $z_s=40\,{a_0}$, and $d_z=10\,{a_0}$ for 100\,${a_0}$ of vacuum, and $W=20\,E_\mathrm{h}$, $z_s=63.75\,{a_0}$, and $d_z=11.25\,{a_0}$ for 150\,${a_0}$ of vacuum.
With these parameters for the CAP, the difference of emitted electrons for 100\,${a_0}$ and 150\,${a_0}$ of vacuum is 1.14\,\%. 
The reduced finite size error with a CAP allows using smaller vacuum regions of 100\,${a_0}$ or less, instead of 150\,${a_0}$, reducing the simulation time per iteration from 61.44 core hours to 40.96 core hours, a 33\,\% speedup, when running on ALCF Theta. 
In general, depending on the targeted problem, a careful convergence test of the vacuum size is required for 2D systems. 

\section{Quantum-mechanical projectile: Electron wave packet}

In the previous section and in most of the literature on electronic stopping, the excitation mechanism is described using a \emph{classical} projectile, i.e., a time-dependent Coulomb potential moving at constant velocity.
It is currently unclear to what extent this approximation becomes unreliable for light projectiles such as protons or electrons.
Electrons are particularly small and light-weight compared to protons or heavy-ion projectiles and the electronic wavelength can reach the scale of inter-atomic distances.
Hence, the approximation of using a classical Coulomb potential to describe electron projectiles is expected to be more severe.
The explicit break-down of this approximation is currently not studied well and systematically.

Treating the incident electron fully quantum-mechanically is, hence, a promising alternative.
Following the work by Tsubonoya \emph{et al.}\cite{tsubonoya2014time}, the initial incident electron can be modeled as a Gaussian-shaped wave packet at the start of simulation,
\begin{equation}
\label{eq:wavepacket}
\psi^{\mathrm{WP}}(\mathbf{r},t_{0}) \equiv \left(\frac{1}{\pi d^{2}}\right)^{\frac{3}{4}} \exp \left[-\frac{(\mathbf{r}-\mathbf{b})^{2}}{2 d^{2}}+i \mathbf{k} \cdot \mathbf{r}\right],
\end{equation}
where $d$, $\mathbf{b}$, and $\mathbf{k}$ are the parameters for defining the spread, the center location, and the wave vector of the wave packet, respectively. 
The wave vector $\mathbf{k}$ represents the group velocity of the incident electron and is the single parameter that controls the kinetic energy of the incident electron.
The time-evolution of this wave packet is described by the time-dependent Kohn-Sham equations, Eq.\  \eqref{eq:tdks}, on the same footing as the rest of the system. 
Thus, the time-dependent Kohn-Sham orbitals include all electrons in the target material and the incident electron of the wave packet.
The electron density is then the sum of the electron density of the target material and the electron density of the wave packet,
\begin{equation}
\label{eq:wavepacket_density}
n(\mathbf{r}, t)=\sum_{i=1}^{N / 2}\left|\psi_{i}(\mathbf{r}, t)\right|^{2}+\left|\psi^{\mathrm{WP}}(\mathbf{r}, t)\right|^{2},
\end{equation} 
where $N$ is the number of electrons in the target material.

\begin{figure}
\centering
\includegraphics[width=\columnwidth]{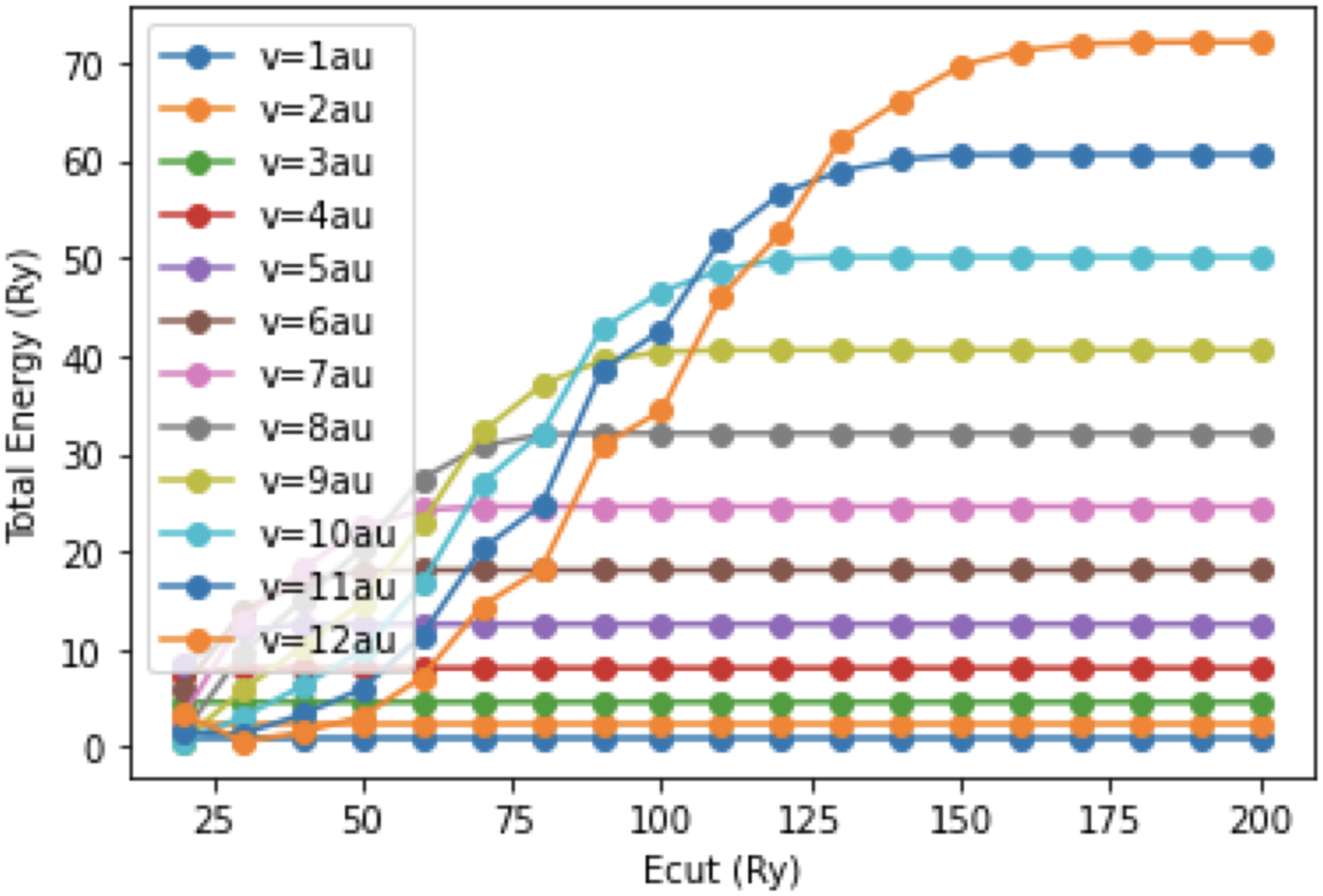}
\caption{\label{fig:ConvergenceTestWP}
Dependence of the total energy of electron wave packets with different kinetic energy moving through vacuum on the plane-wave cutoff used for the simulation.
}
\end{figure}

In the following we characterize the convergence behavior of the Gaussian wave packet with respect to plane-wave cutoff energy (see Fig.\ \ref{fig:ConvergenceTestWP}).
We simulate Gaussian wave packets with different velocities and find that high cutoff values are necessary to converge fast wave packets, possibly leading to a limitation of these simulations.
We also note that the wave packet itself spreads over time, rendering comparison to the classical electron approximation challenging.
Finally, the computation of electronic stopping power $S$ is complicated by the fact that the projectile, if treated quantum mechanically, is part of the electronic system and the approach of computing the stopping power from the increase $dE/dx$ of the electronic total energy is no longer applicable.
Solving this problem remains an open question for future work.

\section{Real-time Electron Dynamics in Aluminum}

\begin{figure}
\centering
\includegraphics[width=\columnwidth]{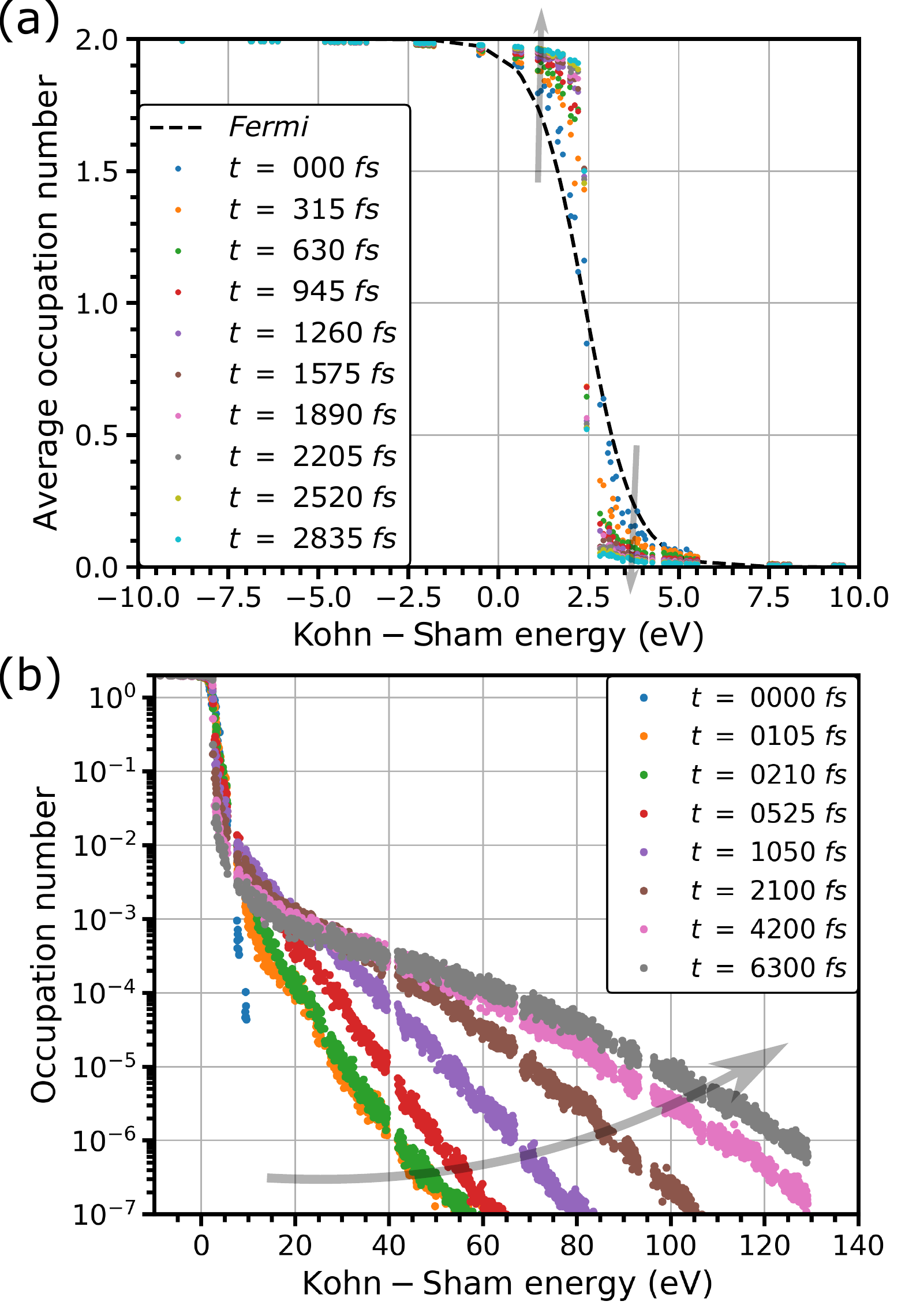}
\caption{\label{fig:ALDA_eq}
Electron dynamics computed for Al with fixed ions using real-time TDDFT.
In (a) we show occupation numbers of the different Kohn-Sham states, averaged over 315 fs of simulation time, as a function of their Kohn-Sham energy.
``000 fs'' shows the average taken from $t$=0 fs to $t$=315 fs etc.
In (b) we plot a single snapshot at shown simulation time for high-energy eigenstates.
Semi-transparent grey arrows guide the eyes for how occupation numbers evolve over time. 
}
\end{figure}

In the following, we explore using real-time TDDFT to simulate electronic thermalization in metals, which is generally assumed to be fast, on the order of 10\,--\,100 fs\cite{Ashcroft_book_1976}.
Previous studies applying the $GW$ method to compute the self energy for Al support this assumption, where the lifetimes mediated by electron-electron scattering are found to be a few tens of fs at energies further away from the Fermi energy and on the order of 100 fs when nearing the Fermi energy\cite{ladstadter:2004,Campillo_PRL_1999,Zhukov_PRB_2005,Schone_PRB_1999}.
Given these short time scales, real-time TDDFT in principle can be used to perform statistical ensemble sampling of an electronic system in internal thermodynamic equilibrium and to calculate expectation values of an observable under different conditions\cite{Modine:2015}.
This is similar to Mermin DFT \cite{mermin1965thermal}, but such a real-time approach can potentially capture additional dynamic effects using the same exchange-correlation functional. 

To this end, Modine \emph{et al.}\cite{Modine:2015} previously explored the idea of performing statistical mechanics on electronic systems, in analogy to simulations of statistical thermodynamics using classical molecular dynamics.
As a first step towards this idea, they initiated a 100\,fs RT-TDDFT simulation using adiabatic LDA for an excited electronic system of Al with fixed ions.
They showed that although the distribution of the time-averaged occupation numbers is Fermi-like, it seems to decrease more sharply near the Fermi energy and takes longer to reach asymptotic values~\cite{Modine:2015}.
To further understand this behavior, we performed significantly longer RT-TDDFT simulations ($>$ 1\,ps) for the same Al system.
We used the same plane-wave cutoff energy of 20\,Ry, $\Gamma$-only Brillouin zone sampling, and the same 32-atom cell. 
In our simulations, this 32-atom cell is either an ideal crystal or a snapshot of a molecular dynamics simulation with a temperature of 7900\,K.

In Fig.\ \ref{fig:ALDA_eq}a we show the resulting long-term electron dynamics in Al with fixed ions, simulated with real-time TDDFT up to $\sim 6$\,ps.
Following the approach by Modine \emph{et al.}\cite{Modine:2015}, the initial wavefunction was prepared in such a way that the distribution of its occupation numbers is close to the Fermi distribution at a given temperature.
For Fig.\ \ref{fig:ALDA_eq} we used 7900\,K and at $t=0$, we can see that the dark blue dots loosely follow the Fermi distribution of the same temperature.
This is, by construction, expected for initial states that are thermal states \cite{Modine:2015}.
We would then expect the occupation numbers to fluctuate around an average that corresponds to this Fermi distribution.

In contrast to this expectation, Fig.\ \ref{fig:ALDA_eq}a clearly shows that the distribution deviates more and more from the initial Fermi distribution as time propagates, indicated by semi-transparent gray arrows.
If we focus on the dynamics near the Fermi energy, the drop in occupations at the Fermi surface becomes steeper and steeper, which is usually associated with lower electronic temperature.
However, the total energy of the system is conserved.
To analyze this further, we also investigate high-energy eigenstates in Fig.\ \ref{fig:ALDA_eq}b, showing that their occupation numbers grow over time, indicating that electrons are promoted to higher energy
states and providing a mechanism for energy conservation.

Since scattering of electrons into higher energy states during electronic thermalization is counter-intuitive, we first thoroughly examine the effect of the initial wavefunction and several numerical parameters.
We ensured that over the simulation time of about 6.3\,ps, the total energy of the system remains conserved within acceptable numerical error of $<0.1$\,meV/atom, suggesting that the numerical time integrator remains stable for the whole simulation.
We also tested that this behavior is independent of the cell size by comparing the dynamics of occupation numbers of high-energy eigenstates in the 32-atom cell to a 108-atom cell, finding again a high-energy tail emerging over time.
Furthermore, we excluded the symmetry of the lattice as a factor by comparing the dynamics for relaxed ($T=0$ K) atomic positions vs.\ a $T=7900$\,K molecular dynamics snapshot.
In addition, we excluded an influence of the particular real-time TDDFT implementation by comparing the Qb@ll and Soccoro \cite{socorro_web} codes.

The occupation number of eigenstate $i$ at simulation time $t$, $f_{i}(t)$, is defined as
\begin{equation}
\label{eq:occ}
f_{i}(t)=\sum_{j=1}\left|\left\langle\phi_{i}|\psi_{j}(t)\right\rangle\right|^{2},
\end{equation}
where the reference states $\phi_{i}$ can be either the DFT ground state or instantaneous adiabatic eigenstates of the time-dependent KS Hamiltonian.
An influence of the reference states used to compute the occupation number was excluded by comparing the adiabatic ground state and the eigenstates of the instantaneous TDKS for projection.
Finally, we also compared different approaches of creating the initial electronic excitation by (i) using the above described thermal state\cite{Modine:2015},  (ii) promoting one electron from valence to conduction band by changing the Kohn-Sham occupation number, and (iii) imposing a vigorous time-dependent displacement of randomly selected atoms.
In all cases we observed the same behavior shown in Fig.\ \ref{fig:ALDA_eq}.

\begin{figure}
\centering
\includegraphics[width=\columnwidth]{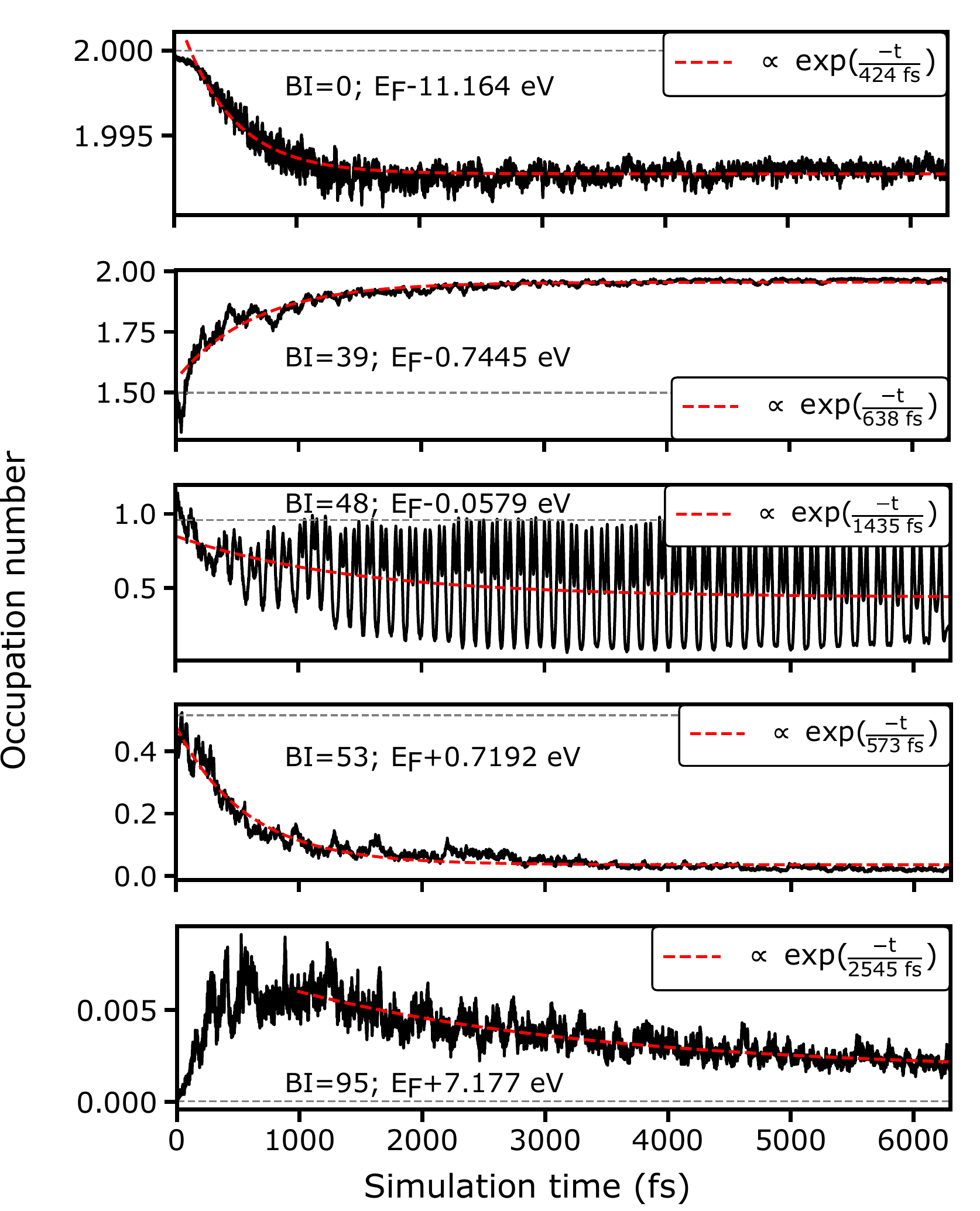}
\caption{\label{fig:occ_time}
Occupation number as a function of simulation time for selected eigenstates. 
These, otherwise randomly chosen, eigenstates have initial occupation numbers of roughly 0.0, 0.5, 1.0, 1.5, and 2.0.
Red dashed curves show the fit against the exponential $a+b\cdot\exp(-t/c)$ to extract the characteristic time scale.
Grey dashed horizontal lines indicate the expected occupation number for a given eigenstate under Fermi distribution. 
Texts describe the band index (BI) and energy difference from Fermi energy at $T=7900$ K.
We found no clear connection of the occupation number dynamics with the energy of the state.
}
\end{figure}

Next, we extract a  characteristic relaxation time by fitting this data to an exponential decay.
We randomly select a few eigenstates across the energy spectrum with initial occupations of about 0.0, 0.5, 1.0, 1.5, and 2.0 and show their dynamics in Fig.\ \ref{fig:occ_time}.
For the following discussion, we refer to them by their band index (BI = 0, 39, 48, 53, and 95, respectively).
First, we notice that they are all evolving away from the occupation number expected based on a Fermi distribution of $T=7900$ K (gray dashed horizontal lines in Fig.\ \ref{fig:occ_time}).
We also notice that the dynamics for the BI=95 state is not monotonic and the occupation number changes in a completely different direction before and after the reflection at around 700 fs.
In addition, fitting the data before 700 fs leads to a characteristic time much shorter than the fit to the data after 700 fs.
Such non-monotonic behavior is not limited to eigenstates with large BI but is commonly observed for other eigenstates.
For these, we only extract the characteristic time for the second part of the dynamics (see the red dashed curve for the BI=95 example in Fig.\ \ref{fig:occ_time}). 
From the extracted characteristic times we found that BI=95, which is far from the Fermi energy, relaxes more slowly than BI=48, which is near the Fermi energy. 
This behavior is different from Fermi liquid theory, which predicts that the lifetime of an eigenstate is longer when its energy is close to the Fermi energy\cite{Baym_book_2008,Campillo_PRL_1999}.
For this reason, and because the excited Al system evolves away from a Fermi distribution, applicability of this relation between Fermi level and lifetime remain unclear.

\begin{figure}
\centering
\includegraphics[width=\columnwidth]{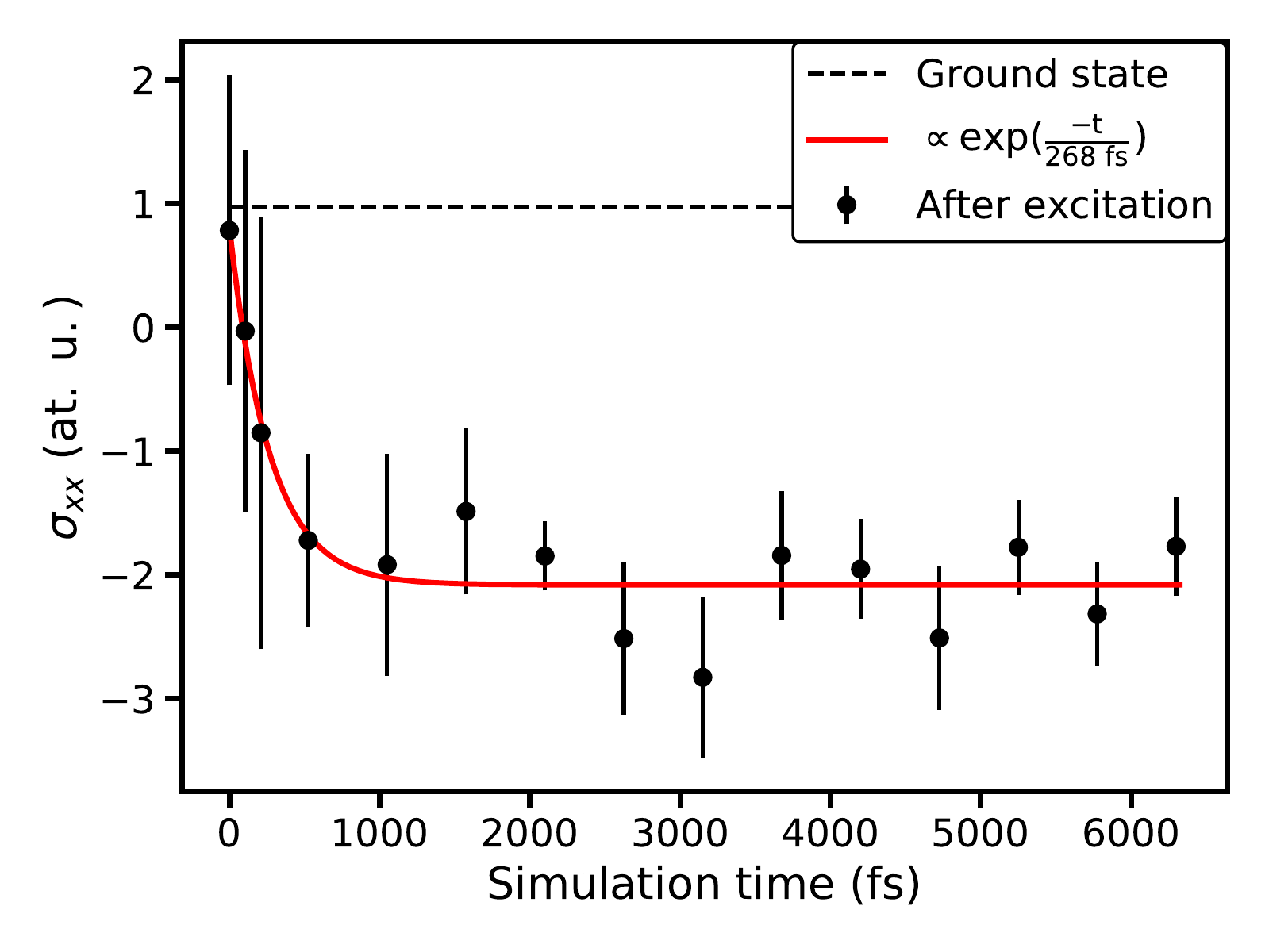}
\caption{\label{fig:stress_dyn}
Time dynamics of the $\sigma_{xx}$ component of the stress tensor, after starting from a thermal state generated with a Fermi temperature of 7900 K.
Stresses are sampled sparsely across the whole simulation and, at each sampled time point, the stress values of the subsequent 10 fs are collected to compute average (solid circles) and standard deviation (error bars). 
The red curve shows a fit against $\sigma_{xx}=a+b\cdot\exp(-t/c)$ to extract the characteristic time scale.
}
\end{figure}

The result is an important, albeit negative, result that points to the inhability of a theory such a TDDFT (at least in its current form) to thermalize electrons.
One potential shortcoming of this analysis may be that from a fundamental point of view, the Kohn-Sham occupation number is not an observable in TDDFT; although that would be an illuminating reason for this inability.

To address this concern, we also analyze stress, which is a functional of the time-dependent electron charge density. 
In Fig.\ \ref{fig:stress_dyn}, we show the real-time dynamics of the stress on the simulation cell after excitation for the $\sigma_{xx}$ component of the stress tensor.
Fitting to this data yields a characteristic time of 268 fs.
The $\sigma_{yy}$ and $\sigma_{zz}$ components have significantly different characteristic times of 889 and 691 fs, respectively, but their dynamics are also monotonic. 
Since a set of independent complex numbers with random phases and magnitudes are drawn from a distribution to construct the initial thermal state\cite{Modine:2015}, the stress and its dynamics are not expected to be isotropic for any given thermal state, but would average out over many thermal states for the same temperature.
We note that these time scales are in the same range as those of dynamics of the eigenstates with monotonic behavior. 
Hence, based on the dynamics of occupation numbers and stress, we conclude that equilibrium is reached over a time scale of 1 ps.
At energies above the Fermi level, $E-E_{F}$, of about 1.0 eV, experimental results report values around 15 fs \cite{Bauer_PSS_2015}.
Computational results include around 30 fs in Ref.\ \onlinecite{Campillo_PRL_1999}, 20 fs from the $GW+T$ method \cite{Zhukov_PRB_2005}, and 70 fs in Ref.\ \onlinecite{ladstadter:2004}.
There is an unresolved discrepancy between Ref.\ \onlinecite{ladstadter:2004} and \onlinecite{Campillo_PRL_1999}, but the literature agrees Al qualitatively follows Fermi liquid theory with band structure effects only giving rise to small quantitative differences.
Not only is this electron-electron relaxation time significantly longer than these results,
but we also find that the system evolves into an unknown distribution with a lower Fermi temperature near the Fermi level and with high-energy tails, compared to the initial Fermi distribution.

\begin{figure}
\centering
\includegraphics[width=\columnwidth]{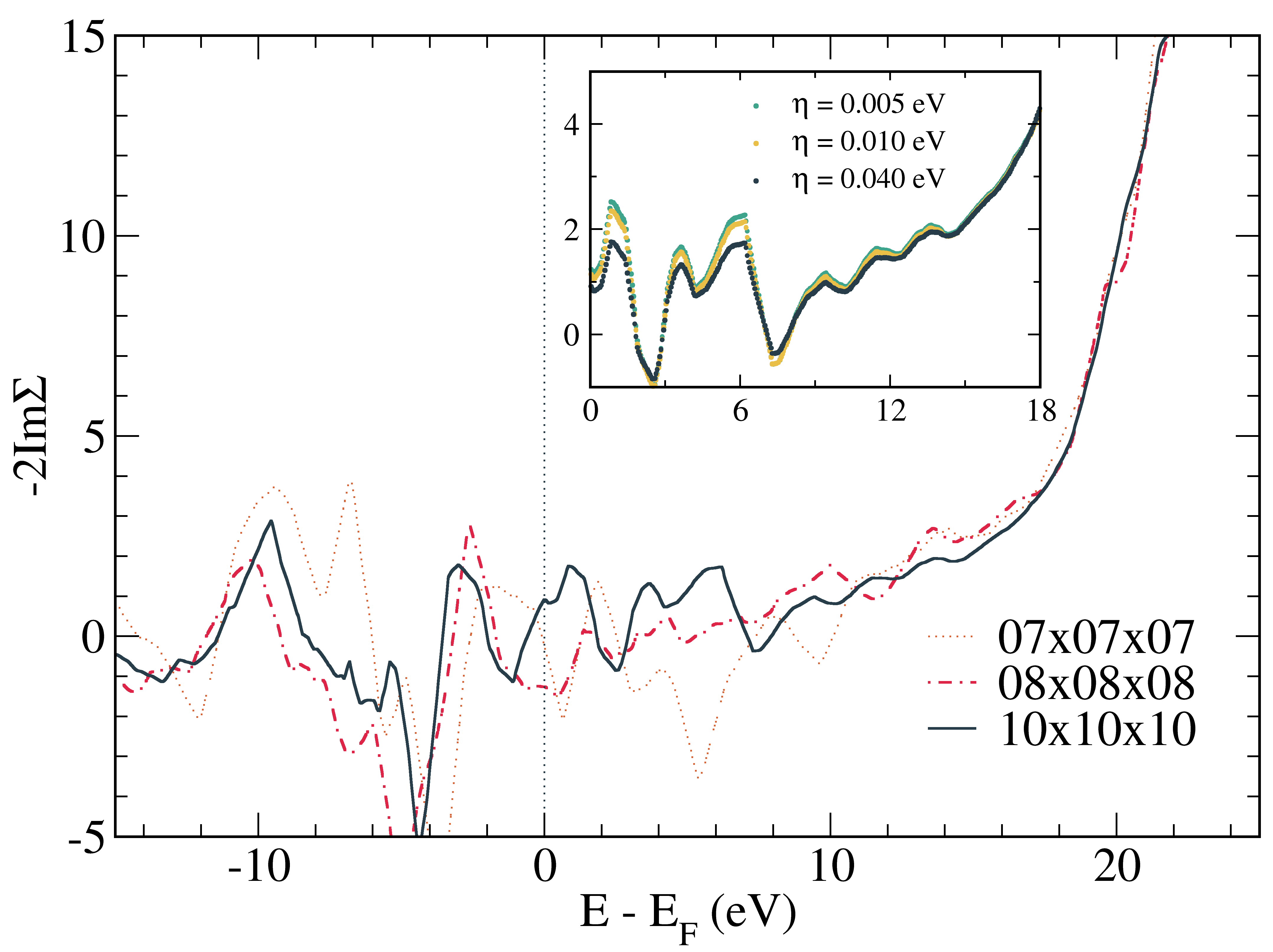}
\caption{\label{fig:Al_-2ImSigma_vs_E-Ef_convergence}
Convergence of $G_0W_0$ calculations with increasing $\mathbf{k}$-point sampling.
The inset shows the results of the 10\,$\times$\,10\,$\times$\,10 $\mathbf{k}$-point calculations with three different $\eta$ values.
The energy range of the inset is the same energy range used for the electron-electron lifetime fit (see text).
}
\end{figure}

Next, we pursue an alternative route to compute the electron-electron scattering lifetime from first principles, based on equating the scattering term to the imaginary part of the electronic self-energy, $\Gamma_{n\textbf{k}}=-2 \operatorname{Im}\left\{\Sigma(\varepsilon_{n\textbf{k}})\right\}/\hbar$\cite{kratzer:2019}.
Computing the imaginary part of the self energy within the $GW$ framework provides lifetimes, using a procedure described by Ladst{\"a}dter \textit{et al.}\ \cite{ladstadter:2004}
Here we use a  computationally more efficient approach by fitting $-2\operatorname{Im} \left\{ \Sigma(\varepsilon_{n\textbf{k}})\right\}$ to a scattering rate of the form $\alpha(\varepsilon_{n\textbf{k}}-E_F)^2$, predicted by Landau's theory of the Fermi liquid \cite{kratzer:2019}.
We compute the imaginary part of the self-energy by performing a $G_0W_0$ calculation where the complex shift $\eta$ of the Kramers-Kronig transformation \cite{kramers:1926,kronig:1926} is set to a value much smaller than what is used in typical $GW$ band structure calculations.
This allows us to accurately resolve the imaginary part of the self-energy near the Fermi energy, see inset of Fig.\ \ref{fig:Al_-2ImSigma_vs_E-Ef_convergence}, and Fig.\ \ref{fig:Al_-2ImSigma_vs_E-Ef_convergence} also illustrates $\mathbf{k}$-point grid convergence tests of our $G_0W_0$ calculations.

\begin{figure}
\centering
\includegraphics[width=\columnwidth]{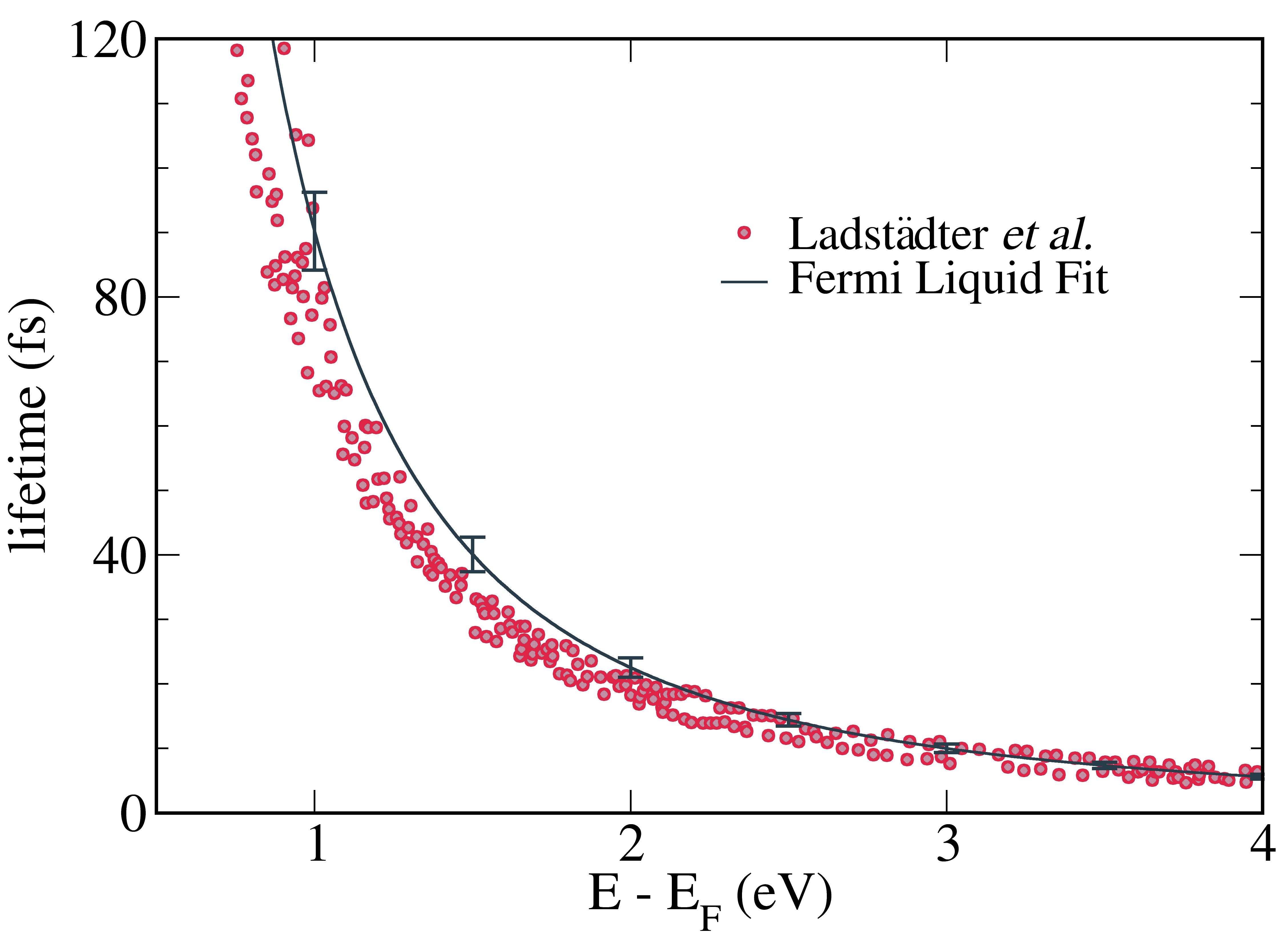}
\caption{\label{fig:Al_ee_lifetime_vs_E-_Ef}
Electron-electron lifetimes obtained from the fit to Landau's theory of the Fermi liquid for the first conduction band at the $\Gamma$ point, computed using a 10\,$\times$\,10\,$\times$\,10 $\mathbf{k}$-point grid and $\eta=0.005$ eV.
Data points were calculated by Ladst{\"a}dter \textit{et al.}\cite{ladstadter:2004}
The error bars show the standard deviation for relaxation times from different $\mathbf{k}$-point grids and $\eta$ values.
}
\end{figure}

Next, we fit the $-2\operatorname{Im} \left\{ \Sigma(\varepsilon_{n\textbf{k}})\right\}$ values for the first conduction band at the $\Gamma$ point, computed using a 10\,$\times$\,10\,$\times$\,10 $\mathbf{k}$-point grid and the smallest value of $\eta=0.005$ eV, over an energy range between 0 and 18 eV to the form from Landau's theory of the Fermi liquid.
The value of $\alpha$ from this fit gives the hot electron lifetimes as
\begin{equation}
\tau_{n\textbf{k}}=\frac{59\textnormal{ fs\  eV}^2}{(\varepsilon_{n\textbf{k}}-E_F)^2},
    \label{eq:tau_nk}
\end{equation}
and is plotted in Fig.\ \ref{fig:Al_ee_lifetime_vs_E-_Ef}.
We include standard deviation error bars at integer and half-integer energy values which compares the lifetimes of the 10\,$\times$\,10\,$\times$\,10 and $\eta=0.005$ eV case to the lifetimes computed from 8\,$\times$\,8\,$\times$\,8 $\mathbf{k}$-point grids with $\eta$ values of 0.005, 0.01, and 0.04 eV and lifetimes from 10\,$\times$\,10\,$\times$\,10 $\mathbf{k}$-point grids with $\eta$ values of 0.01 and 0.04 eV.
The average of the $\alpha$ values from this set of calculations was computed to be 0.0116 (eV)$^{-1}$.
We are satisfied with the use of a 10\,$\times$\,10\,$\times$\,10 $\mathbf{k}$-point grid and $\eta=0.005$ eV due to the error bars being small and the relative error of $\alpha$ being 3.4$\%$ when compared to the average $\alpha$ value.
Figure \ref{fig:Al_ee_lifetime_vs_E-_Ef} shows that our calculated electron-electron lifetimes from the Fermi liquid fit match the lifetimes from the full $GW$ method \cite{ladstadter:2004} well, justifying the future use of this method.
In particular, we note that this approach reduces the computational cost compared to full $GW$ simulations, possibly extending its range of applicability into the high-excitation or warm dense matter regime.
Our calculation of the electron-electron lifetimes predicts that electrons located close to the Fermi energy have lifetimes that are on the order of a few hundred femtoseconds and larger.
For electrons at energies further away from the Fermi energy, our calculation predicts smaller lifetimes on the order of tens of femtoseconds and smaller. 

The relaxation times from the \emph{GW} electronic self energy are about one order of magnitude smaller than our results from TDDFT.
Since we have excluded numerical convergence parameters and finite size effects as possible reasons, we tentatively attribute the unexpected behavior observed in our TDDFT simulations to the limitations of ALDA, which is local in time and space.
The limitations of ALDA for electron-electron scattering were studied before for 1-D model systems\cite{Lacombe:2018,Suzuki:2017} and ALDA is expected to underestimate the scattering probability.
In addition, even the ``exact'' adiabatic functional lacks the ``peak and valley'' features observed in truly exact exchange-correlation potentials and gives rise to spurious oscillations in charge density\cite{Lacombe:2018,Suzuki:2017}.
More generally an explanation for the lack of electron-electron thermalization could be related to the lack of explicit static correlation in the theory, similarly to the problem of electron-\emph{ion} thermalization~\cite{Rizzi2016}.
One could imagine that the promotion of electrons into higher energy states in a 3D metal might be analogous to the charge oscillations observed in the 1-D model.
However, the actual limit of adiabatic semi-local functionals like ALDA remains unclear for condensed systems. 
Future investigation using XC functionals that address self interaction errors (e.g.\ HSE06 \cite{HSE_2003,Krukau_JCP_2006}) or non-adiabatic memory effects (e.g.\ the Vignale-Kohn functional \cite{Vignale_PRL_1996, Wijewardane_PRL_2005}) are needed.
However, such computationally intensive simulations remain impractical at the point of writing.
We also note that other considerations such as choice of pseudopotentials or convergence with respect to Brillouin zone sampling could potentially affect the results to a minor extent.

\section{\label{sec:xc}Exchange and correlation}

Local or semi-local approximations of exchange and correlation are most prevalent in applications of TDDFT to study the dynamics of interacting electrons \cite{Bertsch2000,yabana:2006,Otobe2008,Su2017,Zhang2017,Yamada2019,Goncharov2013,de_giovannini_first-principles_2017,pruneda:2007,schleife:2015,krieger:2015,Pemmaraju2018a}.
This typically means using the adiabatic local-density approximation (ALDA) or its generalized gradient approximation (GGA) extension, but in more recent works\cite{yk_2021_jcp,yk_2019_prl} also modern meta-GGA approximations such as the strongly constrained and appropriately normed (SCAN) functional \cite{SCAN,PW_PP_SCAN} are employed within RT-TDDFT.
More accurate and computationally tractable functionals are always desirable and specifically the influence of long-range corrections, self-interaction errors, and the adiabatic approximation remains unexplored e.g.\ for  electron capture and emission processes.

As a practical approach to move forward, recent progress includes using hybrid exchange-correlation functionals within RT-TDDFT \cite{yabana_2015,Pemmaraju2018a}.
However, their plane-wave implementation carries a computational cost typically about two orders of magnitude higher than that of semi-local functionals \cite{yk_2021_jcp}, rendering  applications to complex extended systems challenging.
The dominant cost of these calculations is the evaluation of exchange integrals. 
To alleviate this problem, some of us pursued the propagation of maximally localized Wannier functions\cite{mlwf_review_2012} in RT-TDDFT, reducing the computational cost of evaluating exact exchange integrals \cite{yk_2021_jcp}. 
Maximally-localized Wannier functions (MLWF) are propagated by \cite{yk_JCP_2019}
\begin{equation}
i \frac{\partial}{\partial t} w_{l}(\mathbf{r}, t) =
\big[\widehat{A}^{ML}
+\widehat{H}[\{w_i\}]
\big]w_{l}(\mathbf{r}, t),
\end{equation}
where the maximal localization operator $\widehat{A}^{ML}$ is an exponential of a unitary matrix that minimizes the spread of the propagating Wannier functions, $\operatorname{min}\left\{\sum_{n}^{N}\left[\left\langle w_{n}\left|\widehat{\mathbf{r}}^{2}\right| w_{n}\right\rangle-\left\langle w_{n}|\widehat{\mathbf { r }}| w_{n}\right\rangle^{2}\right]\right\}_{U}$, and the position operator is $\langle\widehat{\mathbf{r}}\rangle=\frac{L}{2 \pi} \operatorname{Im}\left\{ \ln |\psi| e^{\frac{i 2 \pi}{\mathrm{L}} \widehat{\mathbf{r}}}|\psi\rangle \right\}$. 
For insulating systems with a finite energy gap, the nearsightedness principle of electrons \cite{Prodan11635} allows high spatial localization of time-dependent MLWF orbitals.
This can then be exploited for efficiently implementing hybrid exchange-correlation functionals.
In particular, the spatially localized nature allows to reduce the number of exchange integrals
\begin{equation}
E_{X}=-\frac{1}{2} \sum_{i j}  \iint d \mathbf{r} d \mathbf{r}^{\prime} \frac{w_{i}^{*}(\mathbf{r},t) w_{j}(\mathbf{r},t) w_{j}^{*}\left(\mathbf{r}^{\prime},t\right) w_{i}\left(\mathbf{r}^{\prime},t\right)}{\left|\mathbf{r}-\mathbf{r}^{\prime}\right|}
\end{equation}
that needs to be evaluated.
While time-dependent Kohn-Sham states are generally itinerant, only minimal spatial overlap is expected for distant time-dependent MLWFs and neglecting exchange integrals based on the geometric centers and spreads of the time-dependent MLWFs in the integrand significantly reduces computational cost \cite{yk_2021_jcp}.
\begin{table*}
\caption{
\label{table:hybridXC}
The wall-clock time per iteration for modeling crystalline silicon using a 512-atom simulation cell with the periodic boundary conditions. The planewave cutoff energy of 25 Ry was used with PBE norm-conserving pseudopotentials. ETRS integrator was used with the integration time step of 0.05 at.\,u.
The calculations were performed on 704 processors on 16 Broadwell nodes (Intel Xeon E5-2699A v4 -2.4 GHz) of Dogwood cluster at the University of North Carolina at Chapel Hill. Only MPI (no open-MP/SIMD) was used for this assessment.}
\centering 
\resizebox{0.75\textwidth}{!}{%
\begin{tabular}{l c c c c c c} 
\hline 
& Cutoff distance  & EXX integrals  & Energy drift  & Timer per  & Relative  \\
& ($a_0$) & evaluated (\%) & per iteration ($E_h$) & iteration (s) & iteration time\\
\hline
PBE  & N/A & N/A & $\le$ 1.0 $\times$ $10^{-10}$ & 19.9 & 0.009 \\ 
PBE0 & N/A & 100 & $\le$ 1.0 $\times$ $10^{-10}$  & 2227.8 & 1 \\ 
PBE0 & 25 & 7.4 & 4.1 $\times$ $10^{-7}$ & 271.3 & 0.12 \\ 
PBE0 & 30 & 9.0 & 3.6 $\times$ $10^{-7}$ & 278.4 & 0.13 \\ 
\hline 
\end{tabular}
}%
\end{table*}
Table\ \ref{table:hybridXC} illustrates this reduction of computational cost for a system of 512 crystalline silicon atoms (2048 electrons), when using a cutoff distance for evaluating the exchange integrals needed for the PBE0 hybrid XC approximation\cite{yk_2021_jcp}.
For this test system the computational cost is reduced by an order of magnitude, using a cutoff distance of 25 $a_0$.
We note that due to the nearsightedness principle, this required cutoff distance does not scale with system size.
Consequently, the MLWF approach becomes increasingly appealing for simulations of large systems, because a larger fraction of the exchange integrals can be removed while preserving accuracy.

As an alternative hardware-based paradigm, the high computational efficiency of hybrid XC functional for planewave (RT-TD)DFT codes can be alleviated by adopting GPU architectures.
This is also driven by the growing hybrid CPU/GPU architecture for high-performance computing, aiming to achieve exascale supercomputers\cite{Kothe_CSE_2019}.
Such approach has been successful for ground-state DFT calculations\cite{Hutchinson_CPC_2012,Andrade2013} and RT-TDDFT simulations using parallel transport gauge\cite{Jia_proceedings_2019}.
Andrade \emph{et al.}\ developed a new planewave (TD)DFT code , INQ\cite{Andrade_JCTC_2021}, based on GPU architectures. 
Computationally intensive methods like hybrid functionals are supported in INQ but the speedup remains to be explored in the future.

In terms of how hybrid XC approximations can advance (RT-TD)DFT methodologies, screened range-separated\cite{kronik_2019} and dielectric-dependent hybrid approximations\cite{galli_2019} have emerged as an interesting paradigm in recent years.
Such advanced hybrid XC approcimations could provide an alternative to the computationally expensive many-body perturbation theory framework and potentially enable a description of exciton dynamics in large and complex systems within RT-TDDFT.
Screened range-separated hybrid functionals have been used in linear-response TDDFT to successfully model excitonic features in the absorption spectrum \cite{srsh2019}.
These effects, as well as an accurate description of long-range charge-transfer excitations, typically go beyond standard semilocal approximations for exchange and correlation.
Range-separated hybrid XC approximations are expected to enable a description of charge-transfer dynamics in heterogeneous systems\cite{baer_2009} such as molecule-semiconductor interfaces within RT-TDDFT in combination with the MLWF approach.

While the above-discussed approaches renders hybrid XC functionals more attractive, the computational cost remains significantly higher than for local and semi-local approximations.
Alternatively, we recently demonstrated \cite{sun:2021} the use of a long-range corrected (LRC) kernel \cite{Reining2002} in the context of RT-TDDFT.
The resulting vector potential accounts for the long-range screened electron-hole interaction and is capable of describing excitonic effects in optical  spectra.
At the same time, this RT-TDDFT implementation exhibits computational benefits using massively parallel computing and retains a description of nonlinear effects that are not accessible within the linear response approximation.
We also note that this enables more general future developments around real-time TD current-DFT.

Finally, we note that recent work on the temperature dependence of exchange-correlation models is instructive to consider in working toward a dynamical treatment of thermalization based on TDDFT.
Numerous results have established formal foundations for incorporating electronic temperature in DFT~\cite{pribram2014thermal,pribram2016connection,smith2018warming} and TDDFT~\cite{burke2016exact,pribram2016thermal} beyond the standard Mermin approach~\cite{mermin1965thermal}. 
Building on these foundations, high-quality reference calculations for the uniform electron gas at non-zero temperature~\cite{brown2013path,dornheim2016ab,dornheim2018uniform} have been used to create exchange-correlation functionals~\cite{karasiev2014accurate,karasiev2018nonempirical} and applied to materials in extreme but equilibrated conditions~\cite{karasiev2016importance,karasiev2019exchange}.

However, these results concern electrons that are equilibrated at a fixed temperature, not electrons that are in the process of equilibrating.
Because the thermal contribution to exchange-correlation is typically relatively small, it is reasonable to assume that thermalization through electron-ion scattering can be captured by existing adiabatic functionals. 
However, thermalization through electron-electron scattering will require accounting for physics beyond the adiabatic approximation, which is famously challenging.
We note one potentially promising direction from plasma physics, 
in which a correction accounting for electron-electron scattering beyond a mean-field treatment was proposed as a mechanism to improve agreement with quantum kinetic theory~\cite{desjarlais2017density} for the thermal conductivity of non-degenerate hydrogen plasmas.
Investigations of discrepancies in TDDFT or $GW$ for comparably simple systems might yield insights into deficiencies in these approaches, though extrapolating to degenerate systems would likely be a challenge.

\section{\label{sec:future}Summary and Future Directions}

We discussed various interesting lines of recent development in the context of using real-time time-dependent density functional theory for simulations of electron dynamics on femto- to pico-second time scales.
While our efforts have not yet revealed an integrator that outperforms the enforced time-reversal symmetry method, optimization of the stability region of explicit methods, or incorporation of machine-learning techniques may turn out promising.
Periodic boundary conditions straightforwardly reduce computational cost in particular for finite systems.
Treating the projectile particle quantum mechanically is within reach, albeit expensive, but difficulties around the vanishing distinction of projectile electron and those of the host material require further development efforts.
Based on our detailed simulation results, we conclude that reconciling electron-electron scattering from real-time propagation with many-body perturbation theory will require advances in the description of exchange and correlation.
Finally, such advances seem possible, involving maximally localized Wannier functions or a long-range corrected approach to exchange and correlation.

All of these future developments will undoubtedly be impactful for materials discovery and development and can facilitate the tight integration of electronic excitations and ion dynamics.
Efforts in such directions, including those involving machine learning, are currently underway in many groups worldwide.
Going beyond the scope of this present work are interesting and necessary developments that couple electrons and ions, e.g.\ within Ehrenfest dynamics, or even treat ions quantum-mechanically.
At the same time, such developments in most cases will lead to moderately or significantly increased computational cost.
Taking ongoing developments of modern supercomputing architectures into account, this will require simulation codes which can efficiently benefit from graphics processing units, such as the INQ code \cite{Andrade_JCTC_2021}, the successor to Qb@ll.

\section*{Declarations}

On behalf of all authors, the corresponding author states that there is no conflict of interest.

The data that support the findings of this study are available from the corresponding author, A.S., upon reasonable request.

Funding sources are acknowledge in the following section.

\section*{Acknowledgments}

This material is based upon work supported by the Office of Naval Research (Grant No.\ N00014-18-1-2605) and the National Science Foundation (Grant No.\ OAC-1740219).
A.\ K., B.\ R., and A.\ D.\ B.\ were supported by Sandia's Laboratory Directed Research and
Development (LDRD) Project No.\ 218456 and the US Department of Energy's Science Campaign 1.
Sandia National Laboratories is a multi-mission laboratory managed and operated by National Technology \& Engineering Solutions of Sandia, LLC, a wholly owned subsidiary of Honeywell International, Inc., for the US Department of Energy’s National Nuclear Security Administration under Contract No.\ DE-NA0003525.
X.\ A.\ and A.\ A.\ C. work was performed under the auspices of the US Department of Energy by Lawrence Livermore National Laboratory under Contract DE-AC52-07NA27344.
Y.\ Y.\ and Y.\ K.\ were supported by the National Science Foundation under Award Nos.\ CHE-1954894 and OAC-17402204. 
Support from the IAEA F11020 CRP ``Ion Beam Induced Spatio-temporal Structural Evolution of Materials: Accelerators for a New Technology Era'' is gratefully acknowledged.
This research is partially supported by the NCSA-Inria-ANL-BSC-JSC-Riken-UTK Joint-Laboratory for Extreme Scale Computing (JLESC, \url{https://jlesc.github.io/}).
A.\ Schleife acknowledges support as Mercator Fellow within SFB 1242 at the University Duisburg-Essen. E.\ Constantinescu was supported by DOE Office of Advanced Scientific Computing Research under Contract DE-AC02-06CH11357. 
This work was performed, in part, at the Center for Integrated Nanotechnologies, an Office of Science User Facility operated for the U.S. Department of Energy (DOE) Office of Science by Los Alamos National Laboratory (Contract 89233218CNA000001) and Sandia National Laboratories (Contract DE-NA-0003525)

This research is part of the Blue Waters sustained-petascale computing project, which is supported by the National Science Foundation (awards OCI-0725070 and ACI-1238993) and the state of Illinois.
Blue Waters is a joint effort of the University of Illinois at Urbana-Champaign and its National Center for Supercomputing Applications.
This work made use of the Illinois Campus Cluster, a computing resource that is operated by the Illinois Campus Cluster Program (ICCP) in conjunction with the National Center for Supercomputing Applications (NCSA) and which is supported by funds from the University of Illinois at Urbana-Champaign.
An award of computer time was provided by the Innovative and Novel Computational Impact on Theory and Experiment (INCITE) program.
This research used resources of the Argonne Leadership Computing Facility, which is a DOE Office of Science User Facility supported under Contract DE-AC02-06CH11357.

\end{document}